\documentclass[submission, Phys]{SciPost}

\begin{document}

\begin{center}{\Large \textbf{The Dark Side Properties of Galaxies Requires (viable) Modifications to the Verlinde's Emergent Gravity Theory
}}\end{center}

\begin{center}
Gauri Sharma\textsuperscript{1},
Paolo Salucci\textsuperscript{1*}
\end{center}

\begin{center}
{\bf 1} SISSA,Via Bonomea 265, 34136 Trieste, Italy
\\
* salucci@sissa.it
\end{center}

\begin{center}
\today
\end{center}


\section*{Abstract}
{\bf
Dark Matter is an unknown entity in the Universe. Although several fields of astrophysics \& cosmology are trying to endorse this elusive matter, however, its nature remains an open question. Recently, Verlinde\cite{verlinde2017emergent} has proposed the Emergent Gravity theory (EGT), which is creating severe issues for DM identity.  In this work, we have examined the EGT in the light of the kinematics of the spiral and elliptical galaxies.  Results show that the EGT predictions are in good agreement with latter, though some discrepancy appears in the former.  This current work calls for refinement in EGT. 
}

\vspace{10pt}
\noindent\rule{\textwidth}{1pt}
\tableofcontents\thispagestyle{fancy}
\noindent\rule{\textwidth}{1pt}
\vspace{10pt}

\section{Introduction}
\label{sec:intro}
In the late 1980's, Rubin et al.\cite{rubin1980rotational} and A. Bosma\cite{bosma198121} published the observational evidence of non-Keplerian Rotation Curves (RCs) of spiral galaxies; these findings have revolutionized the field of Astronomy \& Astrophysics as well as Cosmology, by involving the existence of a dark particle (Dark Matter (DM)) necessarily beyond the standard model of elementary particles. Since then, DM scenario has started rooting itself to unscramble the scrambled Universe. Nowadays, DM is the building block of the current cosmological model so as the formation and evolution of all the structure of the Universe (e.g., Padmanabhan  (1993)\cite{padmanabhan1993structure}; Springel et al. (2005)\cite{springel2005simulations}). However, the nature of DM and its structural properties are still under debate and facing a lack of accurate/positive detection. Even to claim that this phenomenon is due to a new elementary particle, is far to be proven. Thus, it is fair and also needed to investigate the dark matter's actual emergence, existence and compositions from a different perspective. 

  Recently, E. Verlinde\cite{verlinde2017emergent} has proposed a new idea called the Emergent Gravity theory (EGT), according to which- "Dark Energy has some entangled entropy allied with it. Baryonic Matter (BM) displaces the dark energy then, due to the \textit{volume law} contribution to the entropy, an elastic force procreates. This extra elastic force works as a gravitational force on massive objects (observed through baryonic tracers), which we have believed to be Dark Matter."

The EGT is rising as an alternative to the DM scenario. It works under a very narrow regime, in which objects are assumed to be isolated, spherically symmetrical, and dynamically relaxed in the Dark Energy (DE) dominated Universe. Despite the narrow regime, EGT has been capable of passing several tests successfully: a) the Tully Fisher relation \cite{verlinde2017emergent}, b) the kinematics of the most massive galaxies \cite{ettori2017dark}, c) the observations of weak lensing in the galaxy clusters  \cite{brouwer2017first}. However, some studies are contradicting the EGT scenario such as the kinematics of the early-type galaxies \cite{tortora2017last}, the radial acceleration relation (RAR) of the spiral galaxies \cite{lelli2017testing}, and the perihelia of solar system planets \cite{hees2017emergent}.

In this paper, we are testing EGT under its regime mentioned above, where it makes robust predictions. In details, we have compared EGT prediction with the following global observational properties of the galaxies, 1) The Universal Rotation Curves (URC) of spiral galaxies obtained by 1000 individual rotation curves and confirmed by other 2000 rotation curves in (\cite{persic1995universal}, \cite{lapi2018precision}) and 2) mass distribution of some elliptical galaxies. 
\\
The layout of this paper is as follows: Section(\ref{sec2}) shows the modeling of galaxy kinematics; in section(\ref{sec3} \& \ref{sec4}) we have tested the EGT model with the spiral and elliptical galaxies; section(\ref{sec5}) contains the discussion \& conclusion. 
\section{Modeling the Kinematics of the Galaxies} \label{sec2}
We can represent the rotation curves of spirals employing the universal rotation curve (URC), pioneered in Rubin et al.\cite{rubin1980rotational} expressed in Persic \& Salucci(1991)\cite{persic1991universal} and set in Salucci et al.(2007)(\cite{persic1995universal, doi}). By adopting the normalised radial coordinate $x\equiv r/R_{opt}$, the RCs of spirals are well described by a universal profile, as a function of $x$ and of $\lambda$, where $\lambda $ is one, a choice, among $M_I$,
the I magnitude, $M_D$, the disk mass and $M_{ vir}$, the halo virial mass\cite{doi}.

\begin{figure}[H]
\begin{centering} \label{urc}
 \includegraphics[width=0.91\textwidth]{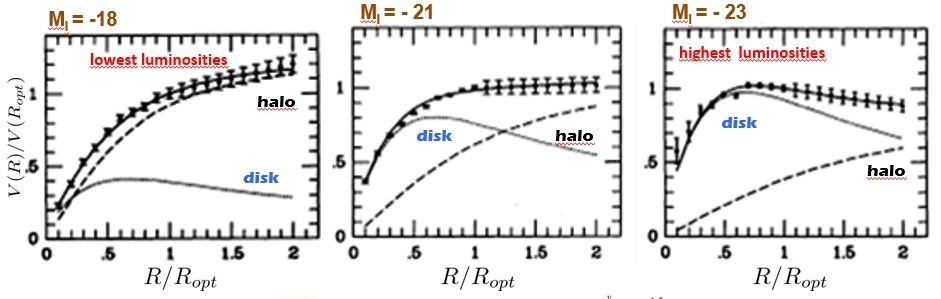}
\caption{URC best fit models of the coadded RCs (points with errorbars) ( Persic et al.(1995)\cite{persic1995universal}). 
It is shown: the corresponding average bin magnitude $M_I$, the  disk/halo contributions (dotted/dashed lines) and
the resulting URC (solid line). $R_{ opt}\equiv 3.2 ~ R_D$.}
\end{centering}
\end{figure}

The universal magnitude-dependent profile is evident in the 11 \emph{coadded} rotation curves $V_{ coadd}(x, M_I)$
(see Fig 6 of \cite{persic1995universal} and Fig(\ref{urc}) in this paper), built from the individual RCs of a sample of 967 spirals with luminosities spanning the whole I-band range: $-16.3 < M_I< -23.4$. I-band surface photometry measurements provided these objects with their stellar disk length scales $R_D$\cite{persic1995rotation}.

The coadded RCs are build as in \cite{persic1995universal}. The URC is the analytical function devised to fit the stacked/coadded RCs $V_{ coadd}(x,M_I)$. In principle, it could be any suitable empirical function of ( $x, M_I$), the idea of \cite {persic1995universal} was that the coadded rotation curves are the averaged circular velocities of spirals,  to choose as the fitting function, the sum in quadrature of the components to the circular velocity. Namely, 1) the stellar disk (with one free parameter); 2) the dark halo with assumed profile with two free parameters, the central density $\rho_0$ and the core radius $r_0$, 3) the HI disk with no free parameter. Then, the data $V_{ coadd}(x,M_I)$ are fitted by the $V_{ URC}$ given in equation(\ref{eq:main}). Where, the first term of the RHS is the stellar disk component, namely the Freeman disk, with the surface density defined as equation(\ref{stellarFreemanDiskDensity}). The stellar disk contribution to the circular velocity is given by equation(\ref{eq2}). The second term is the gaseous disk component, which has been modeled by the Freeman distribution with $R_{HI} = 3R_D$ \cite{evoli2011hi}, therefore, the gaseous surface density scales as equation(\ref{gasFreemanDensity}) and, the contribution to the circular velocity of gaseous disk is given by equation(\ref{eq3}). The third term is Newtonian dark halo of the galaxy defined by equation(\ref{eq4}). Notice that stellar disk mass $M_D$ is a free parameter in modeling and the gaseous mass $M_{HI}$ is given by the empirical relation, explained in \cite{evoli2011hi}.
\begin{equation}\label{eq:main}
V^2_{URC}(x,M_I) \equiv V^2_{URC:d}(x;M_D(M_I)) + V^2_{ URC:HI}(x;M_D(M_I))  + V^2_{URC:h}(x;\rho_0(M_I), r_0(M_I)) 
\end{equation}
\begin{equation}
\label{stellarFreemanDiskDensity}
\Sigma_{D}(R) = \frac{M_D}{2\pi R_D^2} \exp(\frac{-R}{R_D}) 
\end{equation}
\begin{equation}
\label{gasFreemanDensity}
 \Sigma_{HI}(R) = \frac{M_{HI}}{2\pi R_{HI}^2} \exp(\frac{-R}{R_{HI}})
\end{equation}
\begin{equation}
\label{eq2} 
V^2_{ URC:d}(x;M_D(M_I)) = \frac{1}{2} \frac{GM_D}{R_D} (3.2x)^2 [I_0 K_0 - I_1 K_1] 
\end{equation}
\begin{equation}
\label{eq3}
 V^2_{ URC:HI}(x;M_D(M_I)) = \frac{1}{2} \frac{GM_{MHI}}{R_{HI}}(1.1x)^2 [I_0 K_0 - I_1 K_1]
\end{equation}
\begin{equation}
\label{eq4}
 V^2_{URC:h} = \frac{G M_{DM}(R)}{R}
\end{equation}
In EGT scenario, we do not have a dark matter component but Gravity creates an apparent dark halo from the distribution of the \textit{baryonic matter}. Such that, apparent contribution to the circular velocity is: 
 \begin{equation}\label{eq5}
(V^{app}_{DM}(R))^2 =  \sqrt{\frac{G a_0}{6} \\\ \frac{d}{dR}(RM_b(R))}
\end{equation}
and therefore:
\begin{equation}\label{eq6}
(M^{app}_{DM}(R))^2 = \frac{a_0 R^2}{6G} \\\ \frac{d}{dR}(RM_b(R))
\end{equation}
where $a_0$ is an acceleration, having the value $\simeq 7\times10^{-8}cm/sec^2$ and $M_b(R)$ is the total baryonic mass, i.e., the sum of stellar and gaseous mass: 
\begin{equation}\label{eq7}
M_b(R) = M_D(R) + 1.33\footnote{The factor 1.33 is to take into account the He abundance.}\ M_{HI}(R)
\end{equation}
By means the previous equations we can directly calculate the RCs predicted under the EGT framework.\\
\section{Testing Spiral Galaxies } \label{sec3}
\begin{figure}[H]
  \begin{center} 
    \includegraphics[width=11.0cm, height=7.7cm]{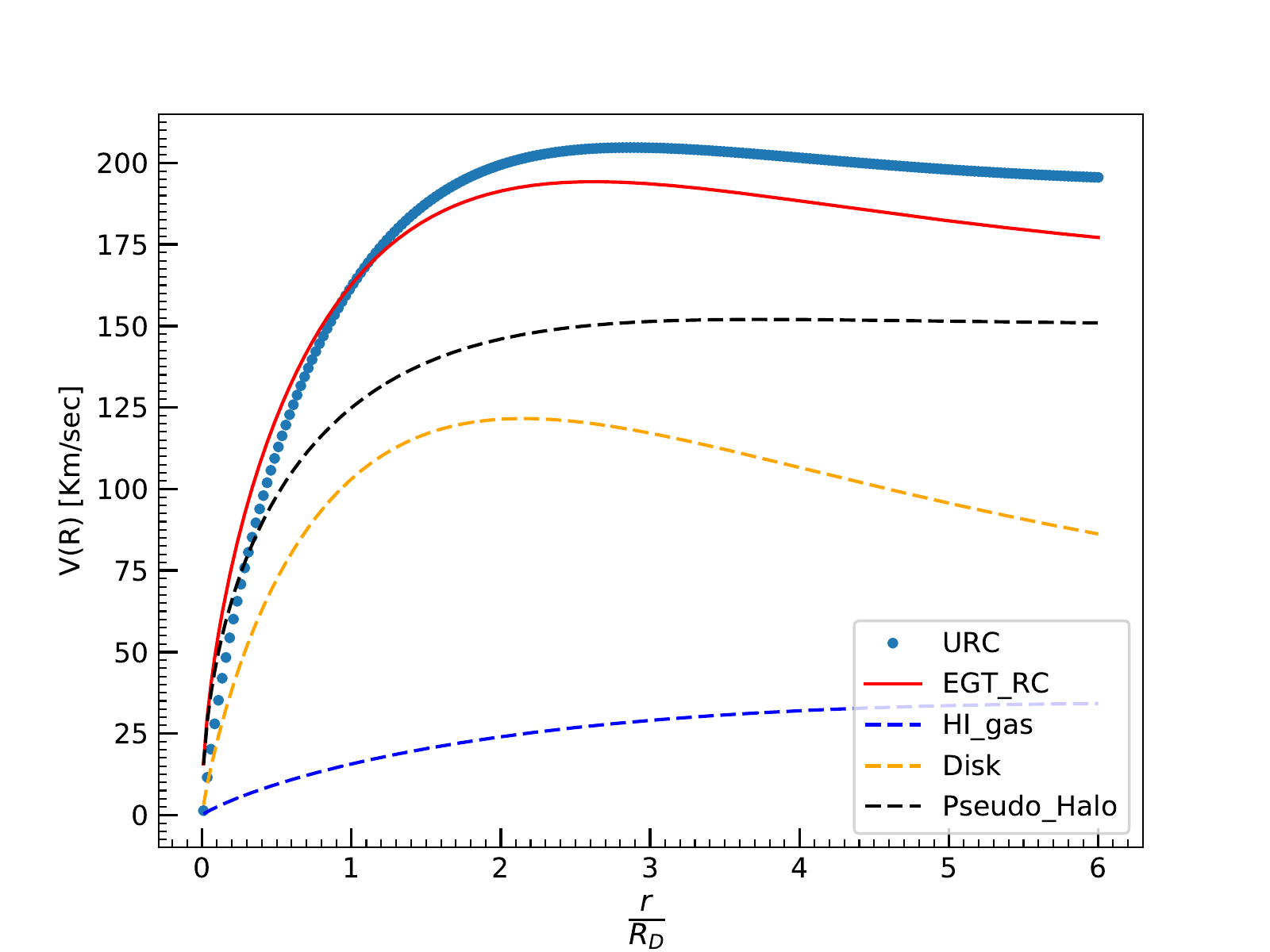}   
    \caption{The URC data (blue dots) for a reference galaxy compared with Rotation Curve predicted under the Emergent Gravity theory (red curve). EGT pseudo halo (black dashed line), stellar disk (orange dashed line) and gas disk (blue dashed line).The circular velocities are normalized at their values on optical ($R_{opt} = 3.2R_D$).}
    \label{fig1}
  \end{center}
\end{figure}
At the first step by employing the URC, we investigate the EG model in the light of a typical rotation curve. In detail, we assume a reference galaxy of the disk length $R_D = 4kpc$ \cite{doi}, disk mass $M_D = 7.8 \times 10^{10}M_\odot$ and viral mass $M_{vir} = 1.3\times 10^{12} M_\odot$. Then, we set ourselves in EGT framework; we have assumed the disk mass $2.5 \times 10^{10} M_\odot$ and the HI mass $3.3\times 10^{10} M_\odot$. The latter distribution is irrelevant in Newtonian plus Dark Matter regime but plays an important role in EGT, while in the outer most region, it may become the main component to $M_b(R)$ in prior distribution. For this reference object, results show that the prediction of EGT with invoking only one free parameter,  are in fair agreement with the URC. The resulting disk mass is a factor two smaller than the estimated in Newtonian gravity plus dark matter regime. However, this discrepancy is within the uncertainties with which we estimate the mass of the galactic disk from the luminosity of galaxies (see figure(\ref{fig1}). 

\begin{figure}[H]
  \begin{center}
    \includegraphics[width=15.0cm, height=12.7cm]{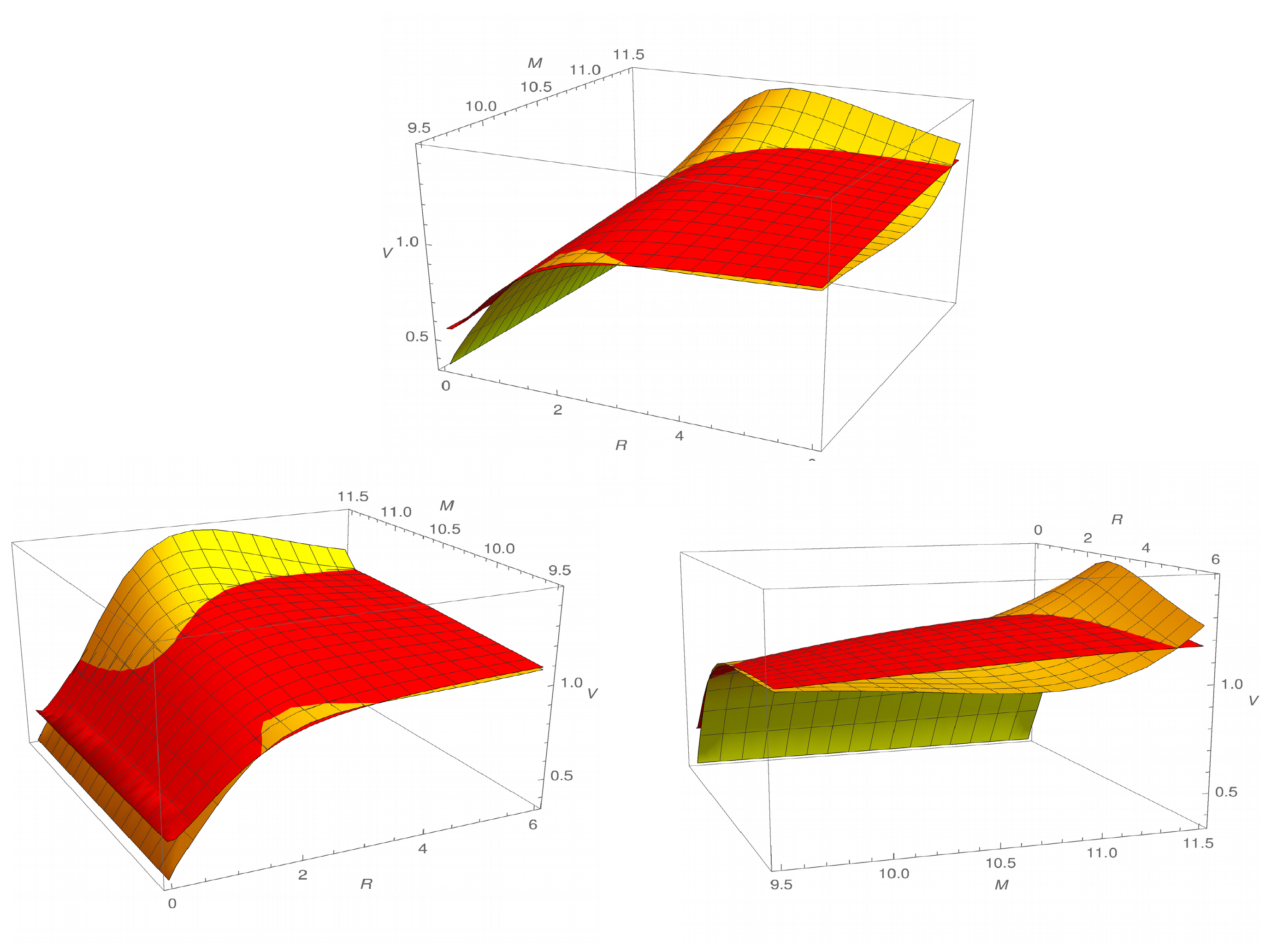}   
    \caption{The URC (yellow) and 3D Rotation Curve predicted under the Emergent Gravity theory (red) and .The velocity ($V$) is  normalized at the optical radius ($R = r/R_D$), M is the logarithmic disk mass in solar units. }
    \label{fig2}
  \end{center}
\end{figure}

This success encourage us to investigate the whole family of spirals by comparing the 3D-URC with RCs emerging from the baryonic matter under the EGT regime given by equation(\ref{eq:main}, \ref{eq2}, \ref{eq3} \& \ref{eq5}). The comparison is shown in figure(\ref{fig2}). We realise that the EGT predictions are a fair representation of actual URC. However, resulting disk mass is factor two smaller than those in URC. There are issues in the comparison at high masses. Although, The disagreement appears in the very inner regions of spirals, however, The overall agreement could consider as surprising. 
In fact, a complex coupling between dark and luminous matter is now substituted quite well only by luminous matter with an exotic interaction, which knows nothing about the alternative Newtonian dark halo of the galaxy.

\section{Testing Elliptical Galaxies}
\label{sec4}
The elliptical galaxies mostly comprise of the old stars and have negligible amount of gas \& dust. They are known as pressure supported systems, such that, to obtain their gravitational potential, one must need a very accurate calculation of the hydro-dynamical equations. However, there are many elliptical galaxies surrounded by hot ionised HI in thermal equilibrium, those allow us to obtain the mass profile from the variation of pressure and the density of the surrounding X-ray emitting gas. In particular, Memola et al.(2009)\cite{memola2009diverse} have built the mass model of two elliptical galaxies NGC7052 and NGC7785. Chandra X-ray Observatory has observed these sources, discussed in \cite{memola2009diverse}. Observations show that- NGC7052 and NGC7785 have an extended x-ray halo out to $16 \ kpc$ and $32 \ kpc$ respectively. The stellar mass distribution is well described by de Vacouleurs spheroid with the half-light radius $R_e = 7.7 \ kpc$ and $R_e = 5.6 \ kpc$ respectively. For the photometric details and the estimate of stellar mass, see Memola et al.(2011)\cite{memola2011dark}.
\\
  
  In EGT, total mass in the elliptical galaxies is the sum of stellar mass $M_{sph}$ and the apparent dark matter $M_{DM}$, defined in equation(\ref{eq11}). Where, we neglect the gaseous mass contribution since the elliptical galaxies have minimal amount of gas. The mass of spheroid $M_{sph}$ is distributed according to de-Vaucouleur profile, in which density distribution is defined by equation(\ref{eq12}). Therefore, the stellar mass distribution as a function of radius $R$ is given by equation(\ref{eq13}).
\begin{equation}
 \label{eq11}
 M_{tot}(R) = M_{sph}(R) + M^{app}_{DM}(R) 
\end{equation}
\begin{equation}
\label{eq12}
\rho_{sph}(R) = \rho_0 \Big(\frac{R}{a}\Big)^{-0.855} \ \exp\Big(\frac{-R}{a}\Big)^{1/4}
\end{equation}
\begin{equation}
\label{eq13}
M_{sph}(R) = \int_0^R 4\pi \rho_{sph}(R) R^2 dR
\end{equation}
where $\rho_{sph}$ is the stellar mass density as a function of radius $R$, $\rho_0$ is the reference density, and $a = \frac{R_e}{b^4}$. where $b$ having the value 7.7. Here, equation(\ref{eq12} \& \ref{eq13}), give the relation between $M_{sph}$ and $\rho_0$. Therefore the EGT mass model of elliptical galaxies takes the form:
\begin{equation}
\label{eq14}
M_{tot}(R) = M_{sph}(R) + \Big(\frac{a_0 R^2}{6G} \frac{d}{dR}(R M_{sph}(R))\Big)^{1/2}
\end{equation} 

\begin{figure}[H]
  \begin{center}
    \includegraphics[width=12.0cm, height=8.0cm]{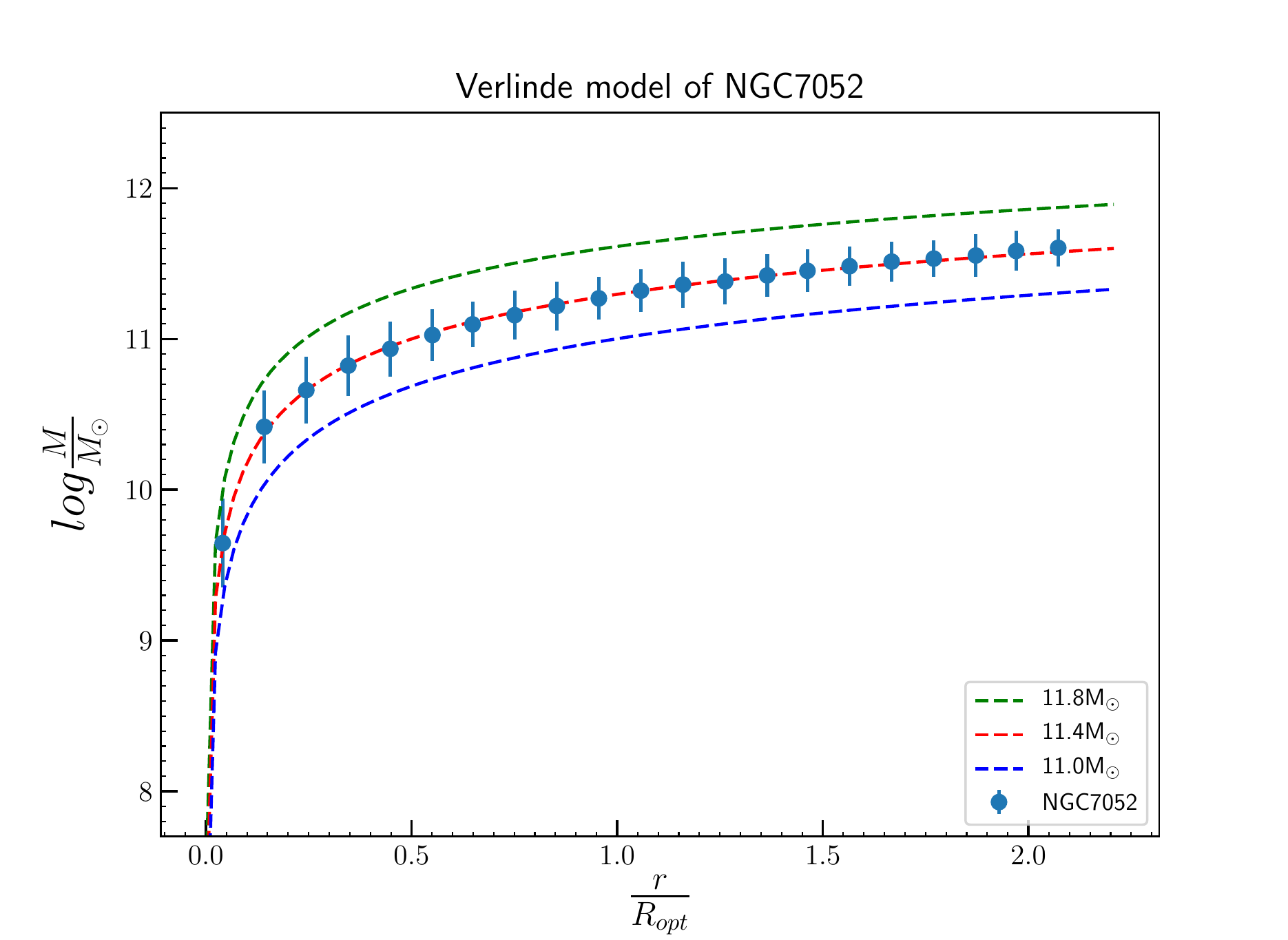}   
    \caption{Mass modeling of NGC7052 under the EGT. Blue dots with error bars is data from Memola et al.(2009)\cite{memola2009diverse} , the red dashed line is the EGT best fit ($M_{sph}=10^{11.4}M_\odot$), green and blue dashed lines are fits with $M_{sph}=10^{11.8}M_\odot$ and $M_{sph}=10^{11.0}M_\odot$ respectively.}
    \label{fig3}
  \end{center}
\end{figure}
\begin{figure}[H]
  \begin{center}
    \includegraphics[width=12.0cm, height=8.0cm]{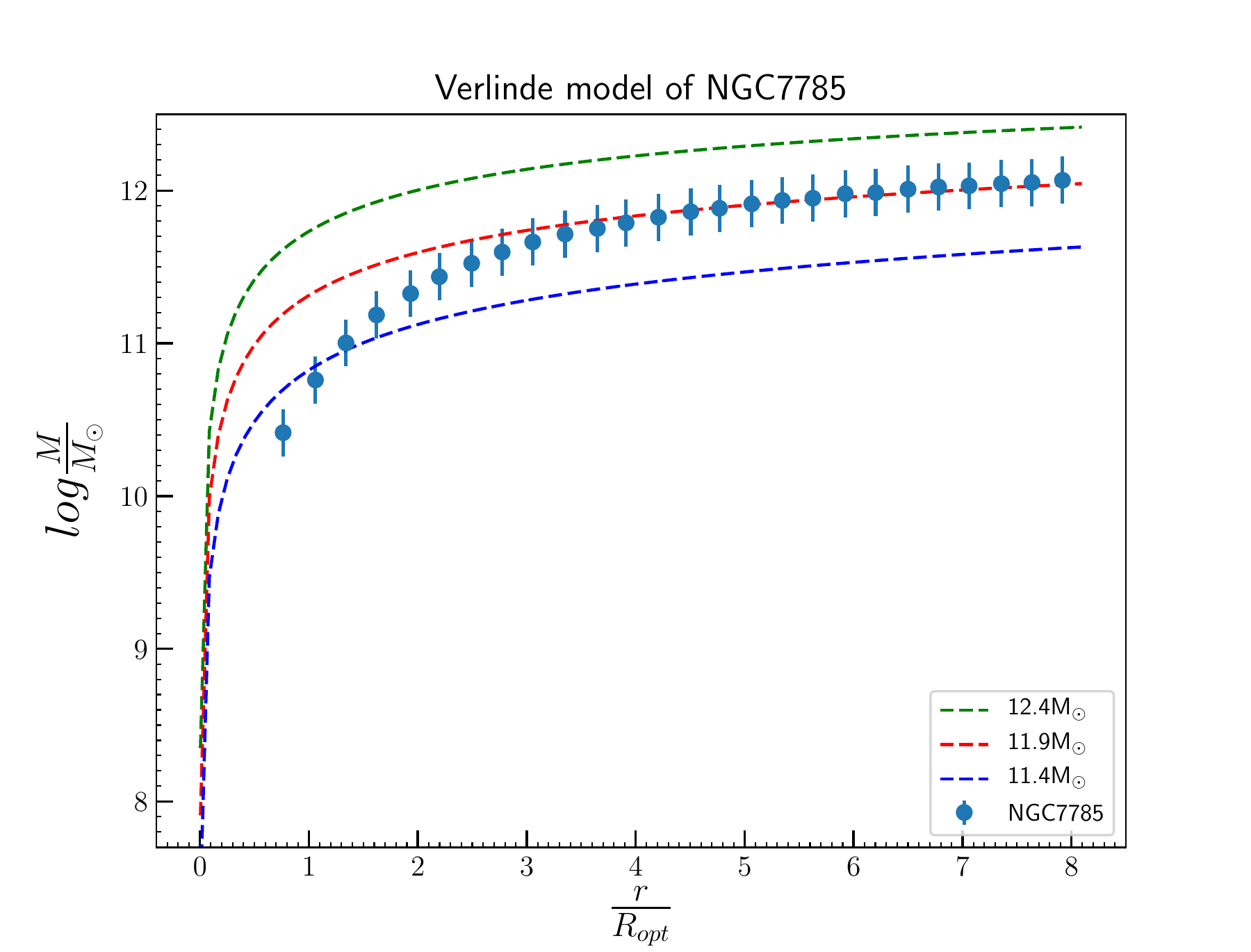}   
    \caption{Mass modeling of NGC7785 under the EGT. Blue dots with error bars is data from Memola et al. (2009)\cite{memola2009diverse}, the red dashed line is the EGT best fit ($M_{sph}=10^{11.9}$), green and blue dashed lines are fit with $M_{sph}=10^{12.4}M_\odot$ and $M_{sph}=10^{11.4}M_\odot$ respectively.}
    \label{fig4}
  \end{center}
\end{figure}

Figure(\ref{fig3} \& \ref{fig4}) shows the EGT mass model of these two elliptical galaxies.  Notice that, this modelling has only one free parameter. The best fit models have: $M_{sph}=10^{11.4}M_\odot$ for NGC7052 and $M_{sph} \sim10^{11.9}M_\odot$ for NGC7785. The value of spheroid mass is similar to the one found in dark matter halo scenario. The fit is excellent for NGC7052 and fairly good for NGC7785. However, concerning the NGC7785, dark halo plus spheroid model have shown the same difficulties in fitting the inner-most region.  

\section{Discussion \& Conclusion}
\label{sec5}
Verlinde\cite{verlinde2017emergent} has proposed a new paradigm to explain the phenomenon of DM, called the Emergent Gravity theory (EGT). Let us rephrase the heart of this theory philosophically, according to this theory-"Universe is the sea of Dark Energy (DE) in which Baryonic Matter (BM) is floating/ traveling. To keep on floating, BM tries to displace the dark energy (likewise ship in the sea). As a consequence, entropy associated with the dark energy procreates an elastic force to support its imbalance (likewise buoyancy force on the ship in the sea). In EGT, this extra elastic force has referred to the Emergent Gravity (apparent DM) which applies to massive objects". All this has consequences on the kinematics of galaxies, which are verifiable. In fact today, it is entirely possible to compare the EGT predictions with a significant amount of kinematical galaxy data. 
\\

  In this work, we have investigated the EGT with the URC obtained for thousands of spiral galaxies \cite{doi}, and the mass model of two elliptical galaxies NGC7052 and NGC7785 from Memola et al.(2009)\cite{memola2009diverse}. Comparison of URC with the RCs driven under EGT finds that the latter are not entirely able to fit the former at high masses and in the innermost regions. Despite that, EGT can reasonably reproduce the URC for low mass spirals at outer radii. Moreover, EGT is in very good agreement with the mass profile of the ellipticals investigated here. Thus, we conclude that the result of this investigation is ambivalent, on one side this new theory reproduces the gross features of the galaxy kinematics without DM, on the other hand, some fine-features are poorly miscarrying.  
\\

  Moreover, reproducing the current observations of DM in galaxies seems not impossible to be obtained by the modified EGT. Let us first say that the EGT predictions are falsifiable, and this theory can reproduce the first order properties of the velocity distribution in galaxies, on the other hand, significant features fail to obtain.  Thus, one should look at the proper refinement of the EGT. To do this, the current results of this work are essential. As shown in comparison, it seems that the discrepancies arise for the massive disk galaxies: does this indicate some galaxy threshold mass in the EGT? Let us draw the attention on few facts: i) EGT works for spherically symmetric systems whereas spirals are not spherically symmetric thus justified to have discrepancies; ii) Apparent Dark Matter (ADM) in EGT allied to entangle entropy of the dark halo, since spirals are not spherically symmetric therefore entanglement of entropy may modify the circular velocity distribution of ADM. Thus, we noticed that in EGT, circular velocity distribution of ADM is no more Newtonian.

\bibliographystyle{SciPost_bibstyle} 
\bibliography{biblio_scipost.bib}

\begin{thebibliography}{10}
\providecommand{\url}[1]{\texttt{#1}}
\providecommand{\urlprefix}{URL }
\expandafter\ifx\csname urlstyle\endcsname\relax
  \providecommand{\doi}[1]{doi:\discretionary{}{}{}#1}\else
  \providecommand{\doi}{doi:\discretionary{}{}{}\begingroup
  \urlstyle{rm}\Url}\fi
\providecommand{\eprint}[2][]{\url{#2}}

\bibitem{verlinde2017emergent}
E.~Verlinde,
\newblock \emph{Emergent gravity and the dark universe},
\newblock SciPost Physics \textbf{2}(3), 016 (2017),
\newblock \doi{10.1086/113063}.

\bibitem{rubin1980rotational}
V.~C. Rubin, W.~K. Ford~Jr and N.~Thonnard,
\newblock \emph{Rotational properties of 21 sc galaxies with a large range of
  luminosities and radii, from ngc 4605/r= 4kpc/to ugc 2885/r= 122 kpc},
\newblock The Astrophysical Journal \textbf{238}, 471 (1980),
\newblock \doi{10.1086/158003}.

\bibitem{bosma198121}
A.~Bosma,
\newblock \emph{21-cm line studies of spiral galaxies. i-observations of the
  galaxies ngc 5033, 3198, 5055, 2841, and 7331},
\newblock The Astronomical Journal \textbf{86}, 1791 (1981),
\newblock \doi{10.1086/113062}.

\bibitem{padmanabhan1993structure}
T.~Padmanabhan,
\newblock \emph{Structure formation in the universe},
\newblock Cambridge university press (1993).

\bibitem{springel2005simulations}
V.~Springel, S.~D. White, A.~Jenkins, C.~S. Frenk, N.~Yoshida, L.~Gao,
  J.~Navarro, R.~Thacker, D.~Croton, J.~Helly \emph{et~al.},
\newblock \emph{Simulations of the formation, evolution and clustering of
  galaxies and quasars},
\newblock nature \textbf{435}(7042), 629 (2005),
\newblock \doi{10.1038/nature03597}.

\bibitem{ettori2017dark}
S.~Ettori, V.~Ghirardini, D.~Eckert, F.~Dubath and E.~Pointecouteau,
\newblock \emph{Dark matter distribution in x-ray luminous galaxy clusters with
  emergent gravity},
\newblock Monthly Notices of the Royal Astronomical Society: Letters
  \textbf{470}(1), L29 (2017),
\newblock \doi{10.1093/mnrasl/slx074}.

\bibitem{brouwer2017first}
M.~M. Brouwer, M.~R. Visser, A.~Dvornik, H.~Hoekstra, K.~Kuijken, E.~A.
  Valentijn, M.~Bilicki, C.~Blake, S.~Brough, H.~Buddelmeijer \emph{et~al.},
\newblock \emph{First test of verlinde's theory of emergent gravity using weak
  gravitational lensing measurements},
\newblock Monthly Notices of the Royal Astronomical Society \textbf{466}(3),
  2547 (2017),
\newblock \doi{10.1093/mnras/stw3192}.

\bibitem{tortora2017last}
C.~Tortora, N.~Napolitano, N.~Roy, M.~Radovich, F.~Getman, L.~Koopmans,
  G.~Verdoes~Kleijn and K.~Kuijken,
\newblock \emph{The last 6 gyr of dark matter assembly in massive galaxies from
  the kilo degree survey},
\newblock Monthly Notices of the Royal Astronomical Society \textbf{473}(1),
  969 (2017),
\newblock \doi{10.1093/mnras/stx2390}.

\bibitem{lelli2017testing}
F.~Lelli, S.~S. McGaugh and J.~M. Schombert,
\newblock \emph{Testing verlinde's emergent gravity with the radial
  acceleration relation},
\newblock Monthly Notices of the Royal Astronomical Society: Letters
  \textbf{468}(1), L68 (2017),
\newblock \doi{10.1093/mnrasl/slx031}.

\bibitem{hees2017emergent}
A.~Hees, B.~Famaey and G.~Bertone,
\newblock \emph{Emergent gravity in galaxies and in the solar system},
\newblock Physical Review D \textbf{95}(6), 064019 (2017),
\newblock \doi{10.1103/PhysRevD.95.064019}.

\bibitem{persic1995universal}
M.~Persic, P.~Salucci and F.~Stel,
\newblock \emph{The universal rotation curve of spiral galaxies: I. the dark
  matter connection},
\newblock arXiv preprint astro-ph/9506004  (1995),
\newblock \doi{10.1093/mnras/278.1.27}.

\bibitem{lapi2018precision}
A.~Lapi, P.~Salucci and L.~Danese,
\newblock \emph{Precision scaling relations for disk galaxies in the local
  universe},
\newblock The Astrophysical Journal \textbf{859}(1), 2 (2018),
\newblock \doi{10.3847/1538-4357/aabf35}.

\bibitem{persic1991universal}
M.~Persic and P.~Salucci,
\newblock \emph{The universal galaxy rotation curve},
\newblock The Astrophysical Journal \textbf{368}, 60 (1991),
\newblock \doi{10.1086/169670}.

\bibitem{doi}
P.~Salucci, A.~Lapi, C.~Tonini, G.~Gentile, I.~Yegorova and U.~Klein,
\newblock \emph{The universal rotation curve of spiral galaxies – ii. the
  dark matter distribution out to the virial radius},
\newblock Monthly Notices of the Royal Astronomical Society \textbf{378}(1), 41
  (2007),
\newblock \doi{10.1111/j.1365-2966.2007.11696.x}.

\bibitem{persic1995rotation}
M.~Persic and P.~Salucci,
\newblock \emph{Rotation curves of 967 spiral galaxies},
\newblock arXiv preprint astro-ph/9502091  (1995),
\newblock \doi{10.1086/192195}.

\bibitem{evoli2011hi}
C.~Evoli, P.~Salucci, A.~Lapi and L.~Danese,
\newblock \emph{The hi content of local late-type galaxies},
\newblock The Astrophysical Journal \textbf{743}(1), 45 (2011),
\newblock \doi{10.1088/0004-637X/743/1/45}.

\bibitem{memola2009diverse}
E.~Memola, G.~Trinchieri, A.~Wolter, P.~Focardi and B.~Kelm,
\newblock \emph{The diverse x-ray properties of four truly isolated elliptical
  galaxies: Ngc 2954, ngc 6172, ngc 7052, and ngc 7785},
\newblock Astronomy \& Astrophysics \textbf{497}(2), 359 (2009),
\newblock \doi{10.1051/0004-6361/200810801}.

\bibitem{memola2011dark}
E.~Memola, P.~Salucci and A.~Babi{\'c},
\newblock \emph{Dark matter halos around isolated ellipticals},
\newblock Astronomy \& Astrophysics \textbf{534}, A50 (2011),
\newblock \doi{10.1051/0004-6361/201015667}.

\end{thebibliography}

\nolinenumbers

\end{document}